\documentclass[a4paper,11pt]{article}
\usepackage{jinstpub} 
\usepackage{lineno}
\usepackage[capitalise]{cleveref}
\usepackage[symbol]{footmisc}

\title{\boldmath New IceTop Trigger in the context of the planned IceCube Surface Detector Enhancement at the South Pole}

\collaboration[c]{on behalf of IceCube collaboration\footnote[1]{\protect\url{http://icecube.wisc.edu}}}

\author{Ek Narayan Paudel}
\affiliation{University of Delaware,\\
104 The Green, SHL 217, Newark, DE, 19716, USA}

\emailAdd{narayan@udel.edu}

\abstract{IceTop is the square kilometer surface array for cosmic-ray air showers of the IceCube Neutrino Observatory at the South Pole. IceTop consists of 81 stations, each comprised of a pair of ice-Cherenkov tanks, which over the years loses sensitivity due to snow coverage. This motivated the plan to enhance IceTop by the deployment of elevated scintillation panels and radio antennas. Coincident detection of an air shower with the IceTop tanks, the scintillators, and the antennas will increase the measurement accuracy of the cosmic-ray properties. While the radio antennas of the enhancement have a higher sensitivity to inclined showers, the current IceTop trigger, requiring coincident hits of both tanks of a station, loses efficiency for such showers. 
Therefore, we studied the feasibility of adding a trigger based on the multiplicity of single tank hits and studied its performance with simulations and data including a one-day test run at the South Pole. In this paper, we present the plans for the surface enhancement and the studies for the new IceTop trigger.}

\begin{document}
\maketitle
\flushbottom

\section{Introduction}
IceTop is the surface cosmic-ray air-shower particle detector of IceCube at the South Pole~\cite{IceCube:2012nn}. It consists of 81 stations with two ice-Cherenkov tanks each. These tanks have digital optical modules (DOMs) with photo-multiplier tubes (PMTs) that can detect Cherenkov photons emitted by charged air-shower particles in the tank ice.

Over the years, there has been continuous accumulation of snow on top of the IceTop tanks resulting into increased attenuation of air-shower particles in the snow and loss in sensitivity.
\begin{figure}[htbp]
\centering
\includegraphics[height=.33\textwidth]{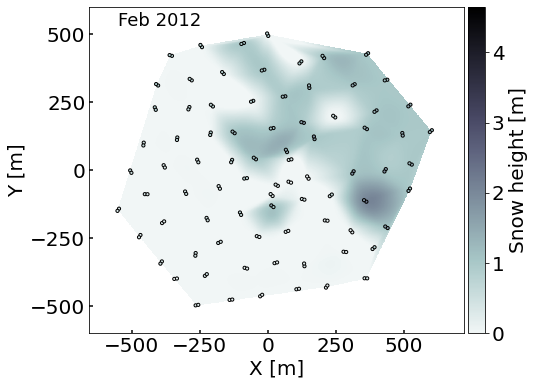}\hspace{-.05\textwidth}
\qquad
\includegraphics[height=.33\textwidth]{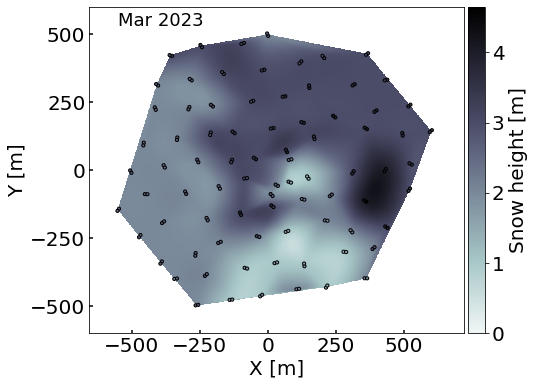} \hspace{-.06\textwidth}
\caption{Snow coverage on top of the IceTop tanks in 2012 (left) and in 2023 (right) as obtained from interpolation of yearly snow height measurements. Pairs of small circles represent IceTop stations (adapted from~\cite{IceCube:2023nyj}).
\label{fig:snow}}
\end{figure}
As can be seen from \cref{fig:snow}, the snow coverage has increased significantly over the last decade and some tanks are now buried about 5 m under the snow. \Cref{tab:ITSMTRun} shows the trigger rate of the standard IceTop trigger for the years 2012 to 2023. The rate was taken from a run in early March of given years. During this period, the standard IceTop trigger rate decreased from $\sim$ 30 Hz in 2012 to $\sim$ 7.6 Hz in 2023.
\begin{table}[htbp]
\centering
\caption{IceTop standard trigger rate (March 2013 -- March 2023)
\label{tab:ITSMTRun}}
\smallskip
\begin{tabular}{|c|c|c|c|c|c|c|c|}
\hline
Year& 2012&2014&2016&2018&2020&2022&2023\\
\hline
Rate [Hz]& 30.2& 24.6&17.4&14.4&11.2&8.5&7.6\\
\hline
\end{tabular}
\end{table}
%



\begin{figure}[htbp]
\centering
\includegraphics[height=.33\textwidth]{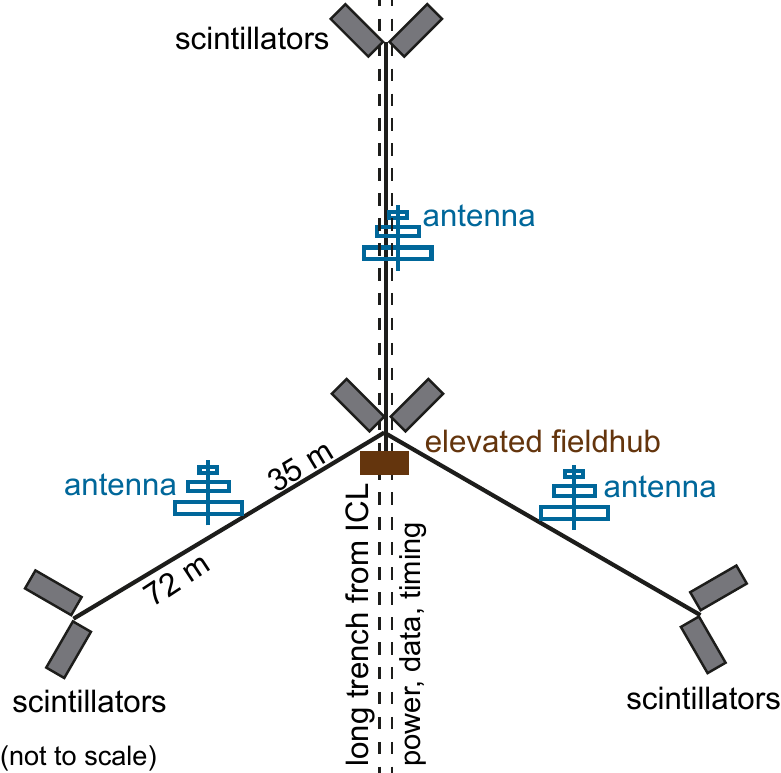}\hspace{-.05\textwidth}
\qquad
\includegraphics[height=.33\textwidth]{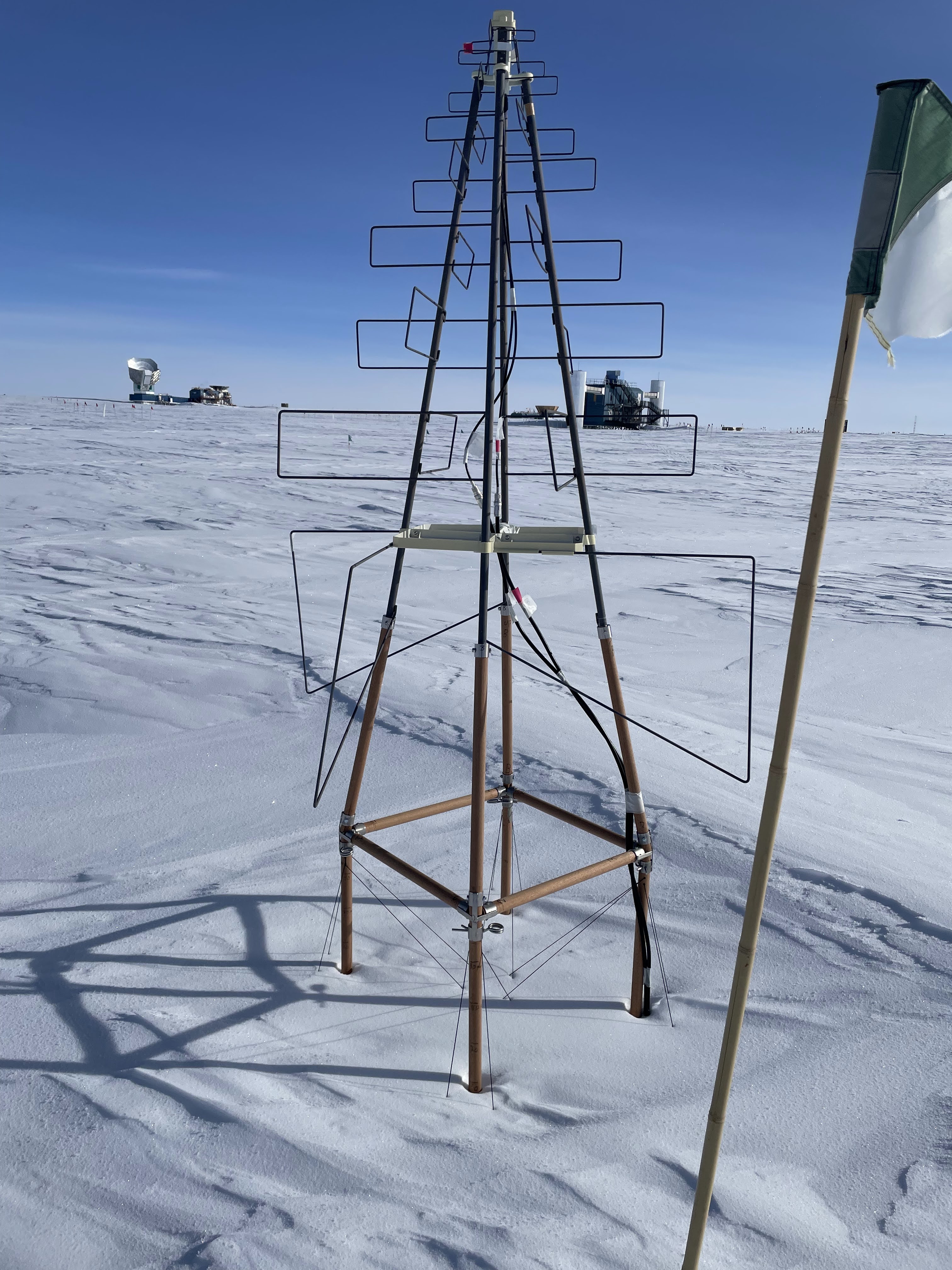} \hspace{-.06\textwidth}
\qquad
\includegraphics[height=.33\textwidth]{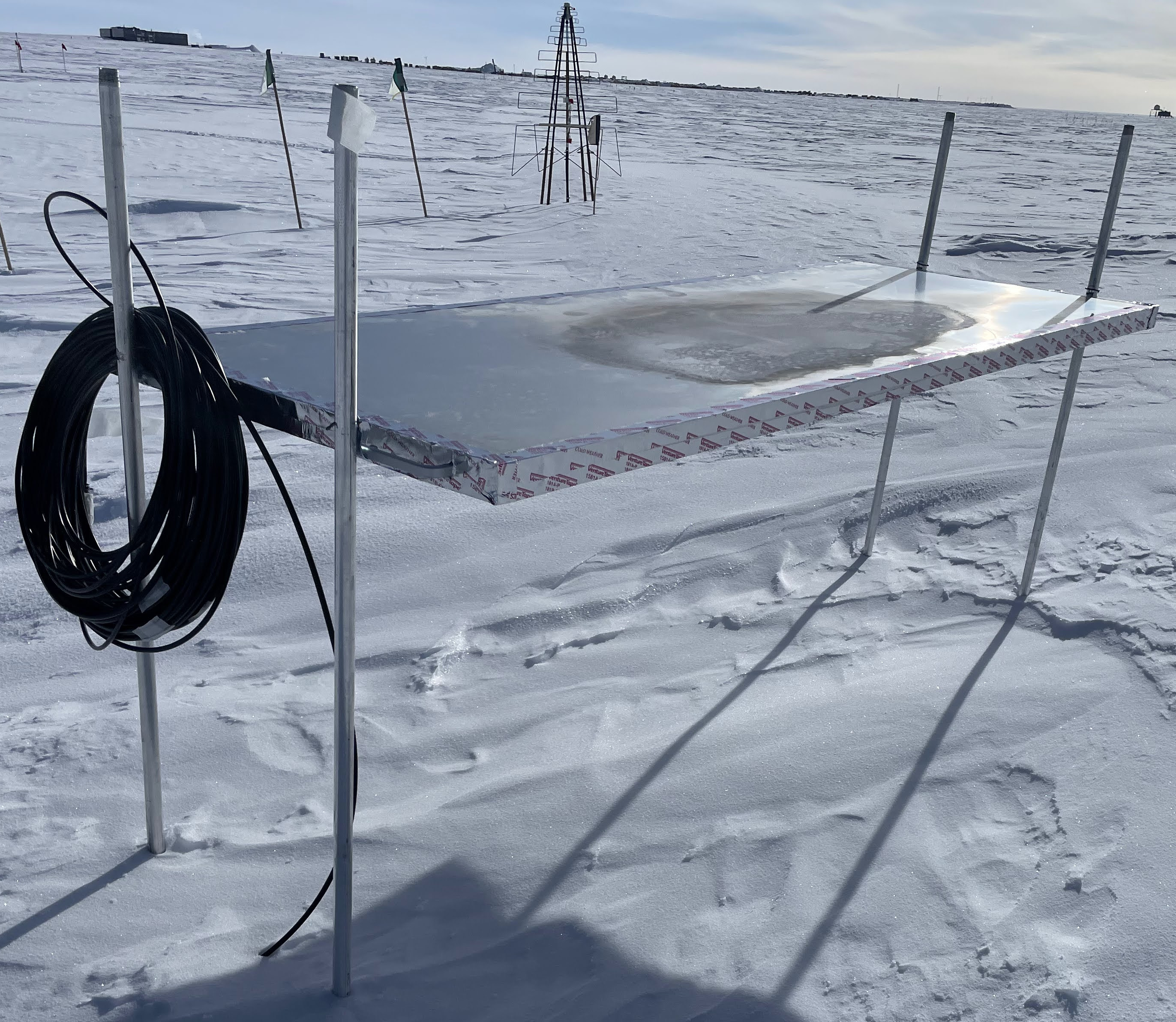}\hspace{.001\textwidth}
\caption{Left: Layout for a single station in the surface array enhancement of IceCube consisting of 4 pairs of scintillator panels and 3 radio antennas; Centre and Right: Radio antenna and scintillator panel (1.5 m$^2$), respectively, which are deployed as a part of the prototype station at the South Pole (Photo credit: R. Turcotte).  \label{fig:proto}}
\end{figure}

%
To overcome the loss in sensitivity due to the effect of snow and also to increase the efficiency of the detector to cosmic-ray air showers of higher inclination angle, there is an ongoing effort to enhance the surface detector with scintillator panels and radio antennas with raisable stands~\cite{IceCube:2023pjc}. Figure ~\ref{fig:proto} shows the layout of a station in the surface array enhancement as well as the images of a radio antenna and a scintillator panel of the prototype station taken in early January, 2023, $\sim$ 3 years after deployment. When needed, these detectors can be raised to keep them above the snow. Additional surface array enhancement stations are already produced and ready to be deployed.

The availability of multiple detectors at the IceTop location allows for better determination of properties of cosmic rays through coincident detection. The main goal of this study is to investigate a new IceTop trigger that can increase the coincident detections of air showers with both radio antennas and the ice-Cherenkov tanks. As the magnetic field at the South Pole is close to vertical ($\sim$ 18$^{\circ}$), inclined cosmic-ray air showers produce stronger radio emission due to the geomagnetic effect and, hence, radio antennas are more sensitive to inclined air showers. For inclined showers, the air-shower particles will be spread over a larger footprint with sparse hits registered in IceTop tanks.
\begin{figure}[htbp]
\centering
\includegraphics[height=.29\textwidth]{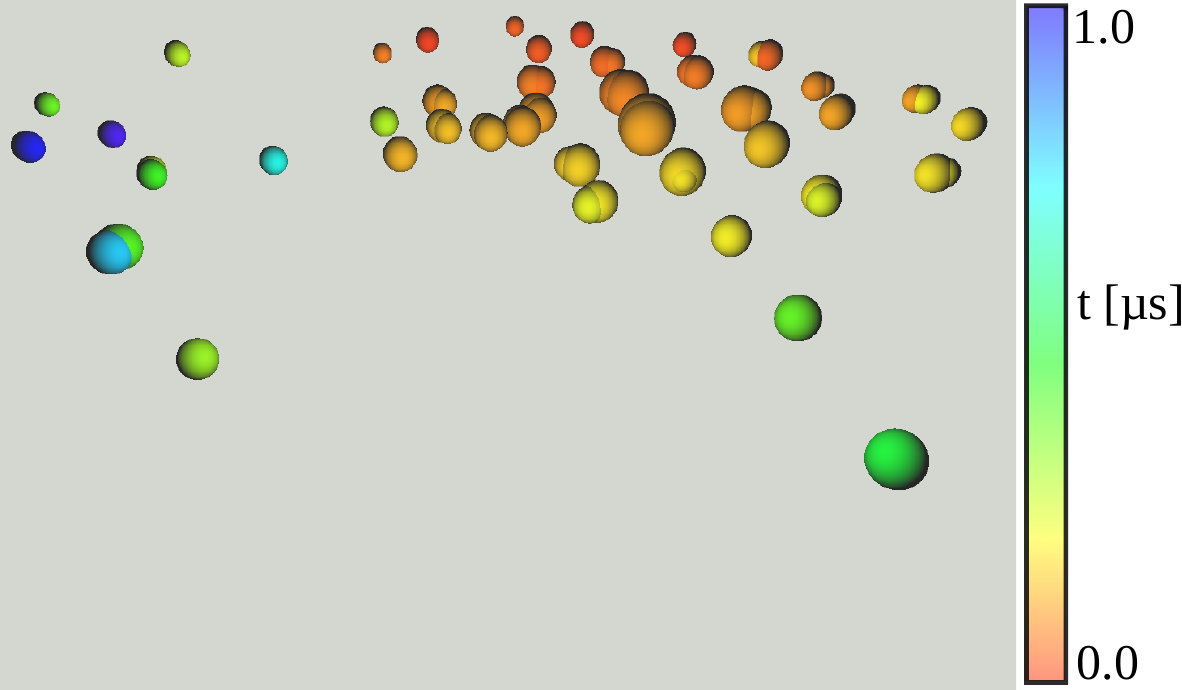}\hspace{-.05\textwidth}
\qquad
\includegraphics[height=.29\textwidth]{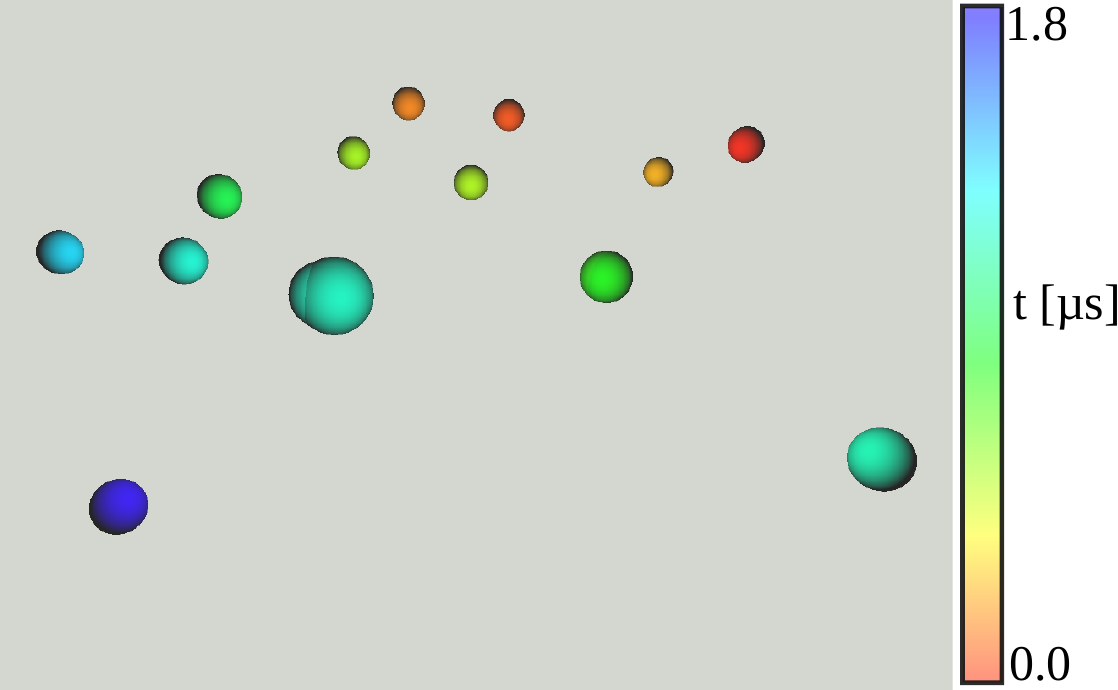}\hspace{.001\textwidth}
\caption{Simulated cosmic-ray air shower at the IceTop surface array. The size of bubbles represents the signal strength of hits and the color represents time, with red indicating early hits and blue representing late hits. Overlapping bubbles appear when both tanks of a station have recorded hits. Left: 10.9 PeV vertical ($\theta \sim$ 7$^\circ$ ) Fe shower with 24 station hits and 18 single tank hits; Right: 10.6 PeV inclined ($\theta \sim$ 57$^\circ$ ) Fe shower with 1 station hit and 12 single tank hits.  \label{fig:steamFe}}
\end{figure}
The event viewer images of air showers in \cref{fig:steamFe} show that an inclined air shower generally has several sparse single tank hits and very few coincident station hits where both tanks within a station register hits.

There is a first level trigger called discriminator trigger in each DOM in the IceTop tanks that initializes digitization of hits that are above a certain threshold. The next level triggers are software triggers that make decisions based on hits recorded in the detector. Previously available data acquisition (DAQ) triggers in IceTop only take coincident station hits into account if both tanks of a station are hit but this becomes inefficient for the showers that are more inclined. So, this study looks for a simple new software trigger for IceTop that will use single tank hits in addition to coincident station hits as an input for the trigger consideration. Each tank has a high gain and a low gain DOM. To prevent double counting of tanks, only hits from high gain DOM within the tanks are used for triggering. We study the trigger threshold and trigger time window that works best for inclined air showers even up to 65$^\circ$ or higher but keeping the trigger rate well below $\sim$ 50 Hz to avoid overloading of the data acquisition system. This relaxation of the trigger requirement will also bring in low energy events of higher elevations, but we plan to develop a filter in the future that will select the inclined events relevant for our study.

\section{Trigger simulation}
As a first step, potential trigger configurations were studied with simulations. Air showers generated using the CORSIKA Monte Carlo simulation code~\cite{Heck:1998vt2} were used for this study with Sibyll2.1~\cite{Fletcher:1994bd} as hadronic interaction model. We used four primaries including p, He, O and Fe with a primary energy ranging from 10$^{14}$ eV to 10$^{17}$ eV and zenith angle $\theta$ $\leq$ 65$^\circ$. This simulation set is not complete as cosmic rays with primary energy above 10$^{17}$ eV are not included but the effect on the trigger efficiency should be negligible. All simulations were weighted using the Gaisser H4a flux model~\cite{Gaisser:2011klf} before calculating simulated trigger efficiency and flux.

The air-shower particles generated by CORSIKA were subjected to detector simulations that included trigger simulation with the experimental IceTop triggers under study. The detector simulation depends on the amount of snow above the IceTop tanks as the shower particles can get attenuated in the snow. The snow height can change over the years and is measured twice every year to keep track of it. We used snow measurements from March, 2021 in our simulations, the most recent ones available at the start of the study. The response of PMTs, DOM electronics, etc., to the Cherenkov photons produced by particles in the ice inside the IceTop tanks is simulated to get the simulated pulse.

\begin{figure}[b]
    \centering
    \includegraphics[width=.9\textwidth]{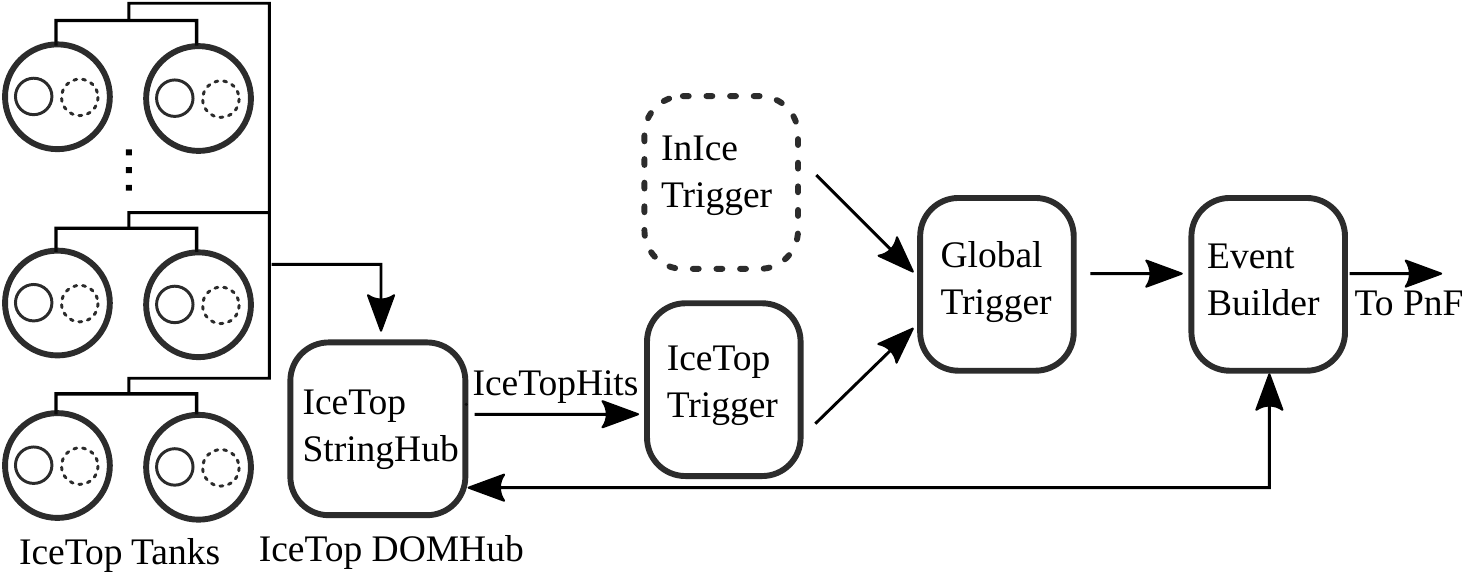}
    \caption{Schematic diagram of a part of the IceCube data acquisition system mainly involved in the IceTop trigger. Each IceTop tank has a high gain DOM (solid circle) and a low gain DOM (dashed circle). The new trigger will be implemented into IceTop trigger modules which will search for triggers in IceTop DOM hits received from the StringHub module. The arrow shows the direction of data flow. PnF is the online processing and filtering system (adapted from Ref.~\cite{IceCube:2012nn}).} 
    \label{fig:daq}
\end{figure}

\Cref{fig:daq} shows the schematic diagram of relevant parts of the IceCube DAQ system. All the hits recorded by IceTop are sent to DOMHub computers in the IceCube Laboratory (ICL). StringHub is the software in DOMHub that handles the data. IceTop hits are sent to the IceTop trigger module for trigger search. The global trigger merges overlapping triggers from various sub-detectors of IceCube and then, based on the timing information from the global trigger, event builder extracts events from the data and sends them to further processing and filtering (PnF)~\cite{IceCube:2012nn}. The experimental IceTop trigger will be able to access single tank hits in addition to coincident station hits. Only hits from high gain DOMs will be used for trigger consideration in the experimental triggers to ensure same tanks are not double counted.

The Monte Carlo simulation does not include background hits from low energy events. To get a realistic and unbiased estimate of the effect on total trigger rate due to background, Fixed Rate Trigger (FRT) events of 2021 measured by IceCube were used. The FRT trigger captures all hits from the detector at a constant rate (every 300 seconds) for a duration of 10 ms. We simulated the experimental triggers on the FRT triggered data to obtain the total rate of triggered events with background for each of the experimental triggers.
\begin{figure}[htbp]
\centering
\includegraphics[width=.6\textwidth]{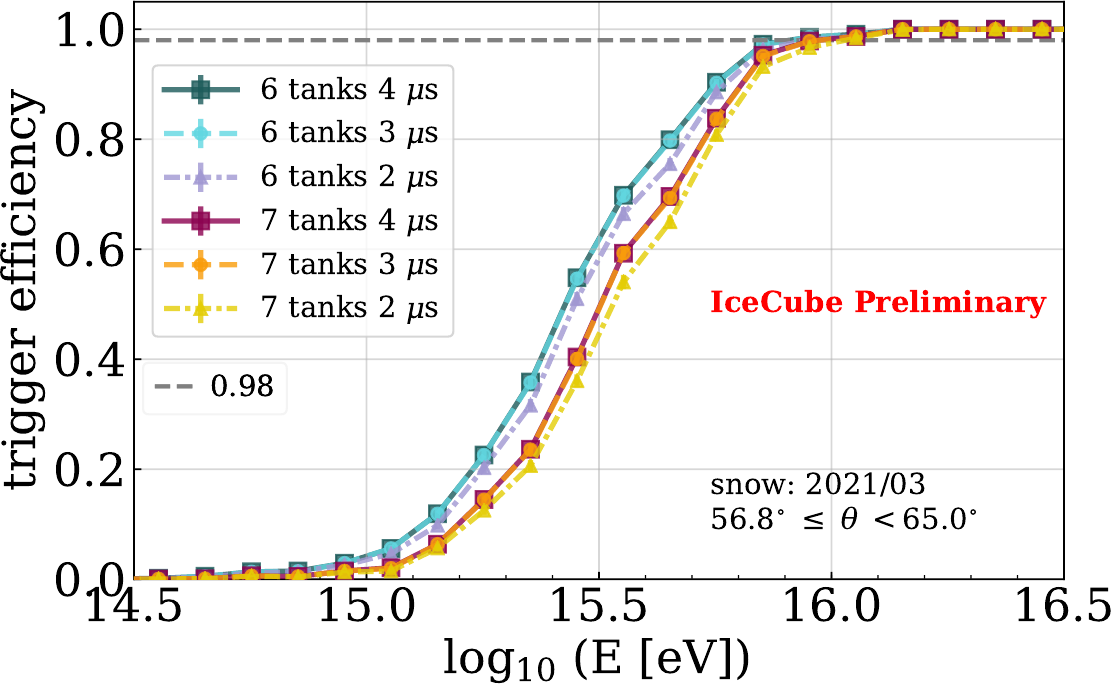}
\hspace{-.06\textwidth}
\qquad
\includegraphics[width=.6\textwidth]{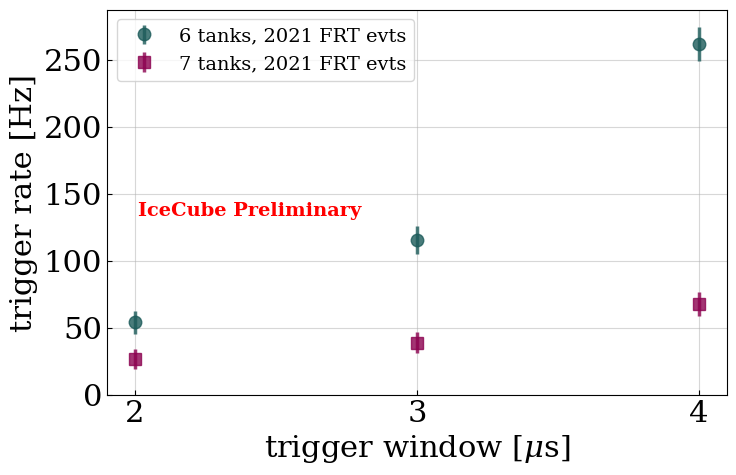}
\caption{Top: Simulated trigger efficiency without background for 6- and 7-fold coincidences of IceTop tanks within coincidence time windows in the range of 2-4 µs over primary energy, for the largest zenith angle bin in the simulated data set. The dotted line represents 98\% trigger efficiency. The trigger efficiency does not improve for coincidence intervals > 3 µs, but the efficiency decreases for smaller intervals. Bottom: Total trigger rate calculated on IceTop data from the Fixed Rate Trigger (FRT) for the year 2021 for 6- and 7-fold coincidences within different time intervals. For a trigger setting of 7 tank hits in 3 µs, the total trigger rate is under the maximum target rate of 50 Hz . Hence, this trigger setting was chosen for additional investigation. 
\label{fig:exptTriggers}}
\end{figure}

The trigger efficiency is defined as follows:
\begin{equation}
\label{eq:trigEff}
\begin{aligned}
\text{trigger efficiency} &= \frac{\text{no. of triggered events}}{\text{total no. of simulated events}}\,.
\end{aligned}
\end{equation}
~\Cref{fig:exptTriggers} shows the simulated trigger efficiency for inclined air showers as a function of logarithm of energy for various experimental triggers with different thresholds and trigger time windows on the top. This only includes physics events as background was not added in the simulation. We have only included air showers for which the shower core is contained within the IceTop footprint and hence has a higher chance of being reconstructed. Trigger efficiency is generally better when requiring only 6 tanks in coincidence instead of 7 tanks for a given trigger window. For both multiplicities, trigger efficiency starts to worsen if the trigger window is reduced to below 3 $\mu$s. The bottom subplot of ~\cref{fig:exptTriggers} shows the trigger rate obtained from FRT data of 2021 plotted as a function of trigger window for different thresholds. This rate would include the effect of any background event seen by the detector. For a trigger window of 3 $\mu$s, the total rate of 6-fold coincidence trigger is well above 100 Hz. In order to keep the additional load to the data acquisition system low, a trigger setting of 7 tanks without any spatial requirements other than the threshold and 3 $\mu$s as the trigger window was selected for further investigation which would have a total rate of $\sim$ 40 Hz for the year 2021 based on FRT data. Simulation shows that only $\sim$ 12 Hz out of the $\sim$ 40 Hz may be from useful cosmic ray events. The rest can be background from low energy events. Hence, a filter will be essential for the usefulness of this trigger.  The name for this trigger is IceTop7HG (IceTop 7 High Gain). As confirmed by the test run described in ~\cref{sec:Test}, we expect these rates to further reduce by the time the trigger is implemented in the 2023 run season due to additional years of snow accumulation on top of IceTop.

\section{Comparison with standard IceTop trigger}
In this section, the performance of the new trigger, IceTop7HG is compared with the standard IceTop trigger called IceTop Simple Majority Trigger (IceTopSMT). Each IceTop station has two tanks and each tank has a high gain (HG) DOM and a low gain (LG) DOM. If both tanks in a IceTop station get hit in coincidence, it is a coincident station hit and if only one tank in a station gets hit, it is a single tank hit. The IceTopSMT trigger generally searches for 3 such coincident station hits within a trigger window of 5 $\mu$s~\cite{IceCube:2012nn}. This requirement of coincident station hits helps to reduce the effect of random single tank hits due to low energy background events. But this requirement makes it less efficient for inclined events which usually feature single tank hits and very few coincident station hits, if any. The IceTop7HG trigger searches for hits from high gain DOMs of any 7 IceTop tanks within the trigger window of 3 $\mu$s. Hence, IceTop7HG can also trigger on inclined air showers with shower particles spread sparsely over a larger footprint. 
\begin{figure}[htbp]
\centering
\includegraphics[width=.6\textwidth]{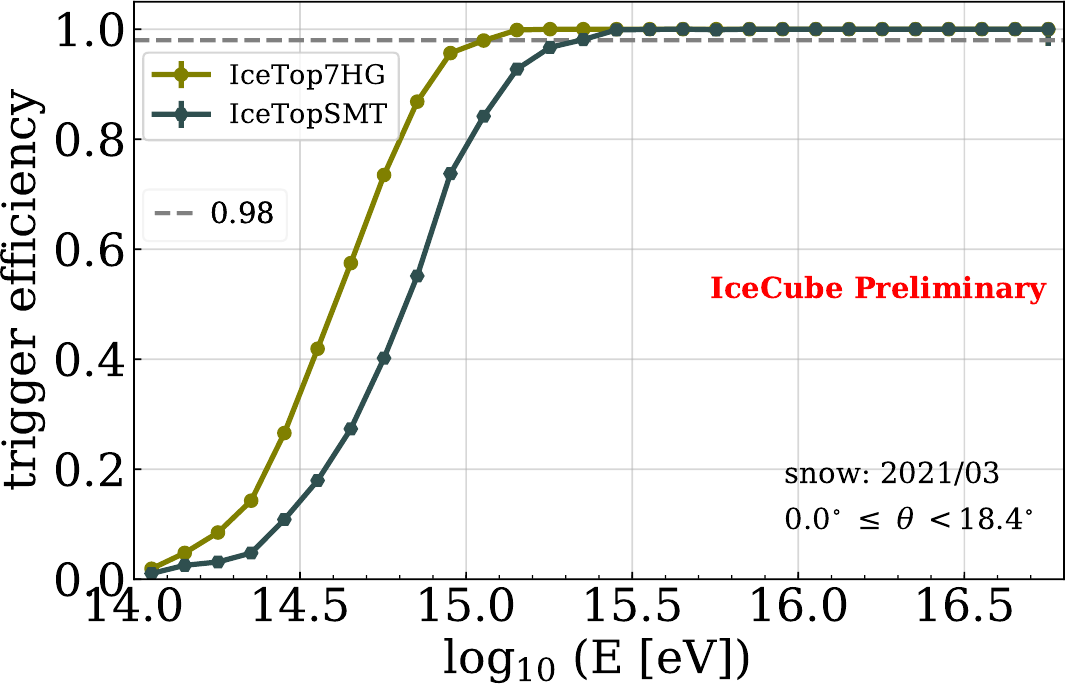}
\qquad
\includegraphics[width=.6\textwidth]{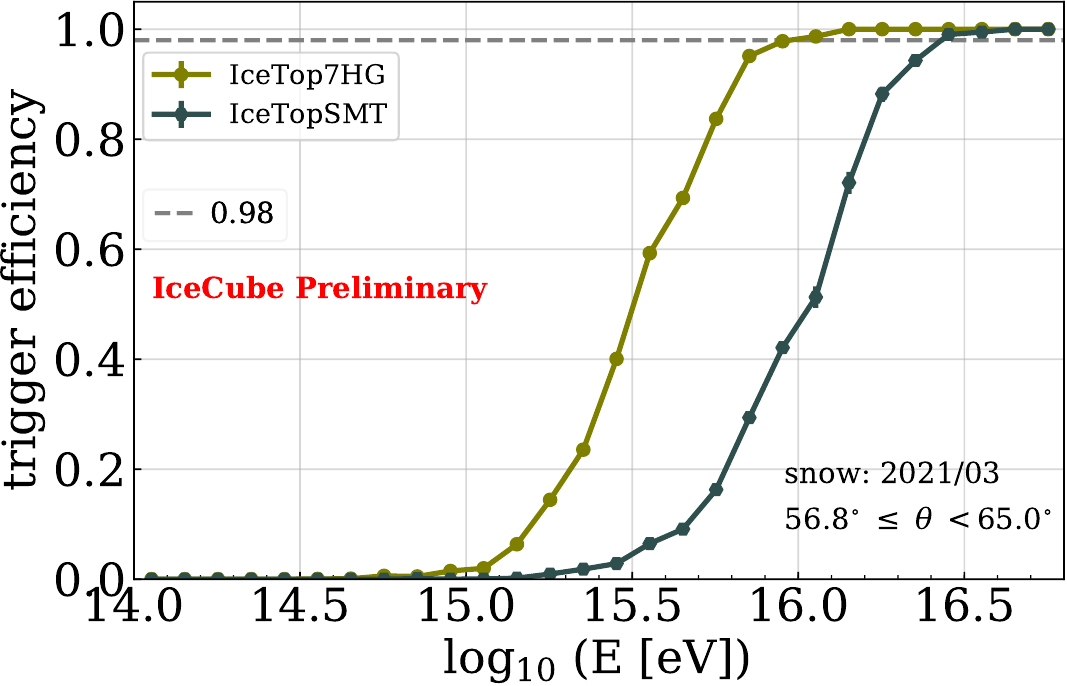}
\caption{Simulated trigger efficiency plotted as a function of primary energy for vertical air showers (0.0$^\circ$ -- 18.4$^\circ$) (top plot) and inclined air showers (56.8$^\circ$ -- 65.0$^\circ$) (bottom plot). The simulation was done with the snow accumulation measured on March, 2021. For, both, vertical and inclined air showers, the IceTop7HG (7 tanks) trigger has a lower energy threshold for full trigger efficiency than the IceTopSMT (3 station) trigger; more so for the inclined case.
\label{fig:selectEff}}
\end{figure}

In ~\cref{fig:selectEff}, it can be seen that for the 2021 snow level, the new trigger IceTop7HG, has a lower energy threshold by about 0.2 in log$_{10}$(E) for full efficiency for vertical showers (0.0$^\circ$ -- 18.4$^\circ$) in comparison to the standard IceTopSMT trigger. The improvement is even better for the inclined zenith bin of 56.8$^\circ$ -- 65.0$^\circ$ where the full efficiency threshold reduces from $\sim$ 30 PeV to $\sim$ 10 PeV, i.e., by about 0.5 in log$_{10}$(E).
\begin{figure}[htbp]
\centering
\includegraphics[width=.6\textwidth]{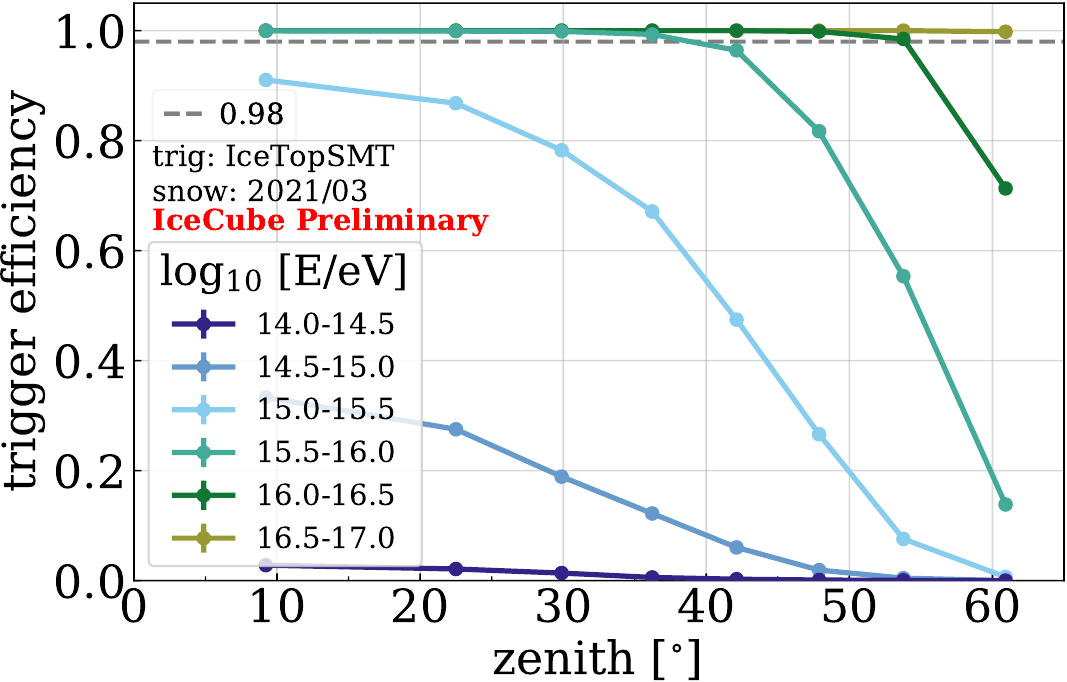}
\qquad
\includegraphics[width=.6\textwidth]{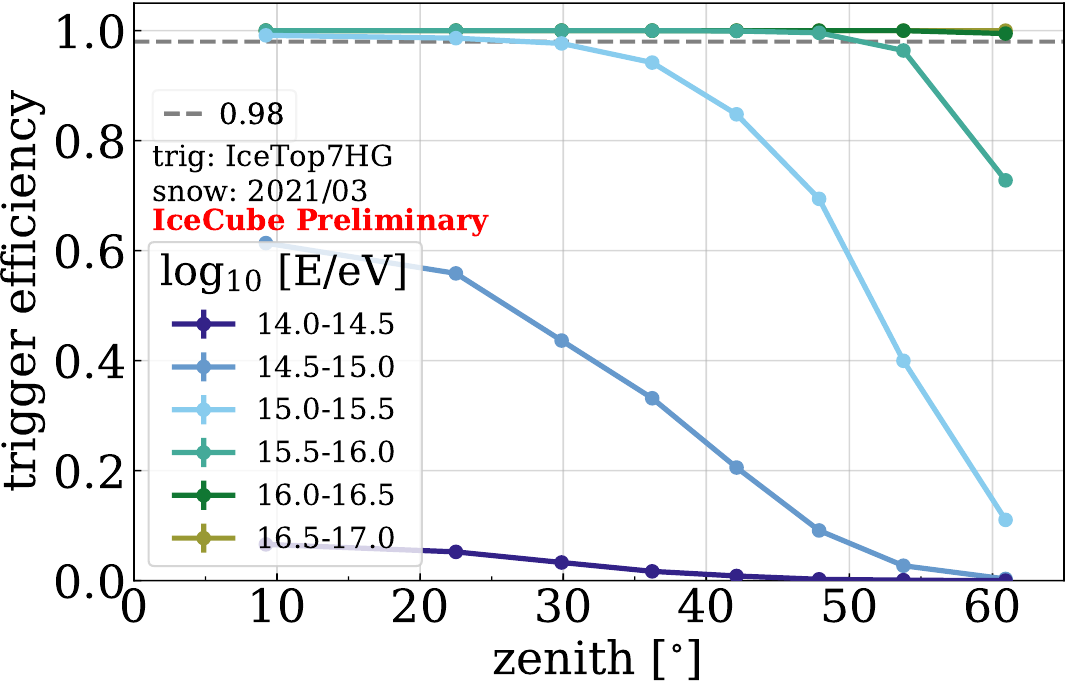}
\caption{Simulated trigger efficiency for the 2021 snow level plotted as a function of zenith angle of cosmic ray primary in various energy bins for the IceTopSMT trigger in the top and the IceTop7HG trigger in the bottom plot.\label{fig:selectEffZenith}}
\end{figure}
When the trigger efficiency is plotted as a function of zenith angles for various energy bins as in \cref{fig:selectEffZenith}, we can see that, both, IceTopSMT and IceTop7HG are never fully efficient for air showers below $\sim$ 1 PeV in energy at all zenith angles. IceTopSMT becomes fully efficient at energies only above $\sim$ 3 PeV  but starts to lose trigger efficiency at larger zenith angle above $\sim$ 40$^\circ$. For energies above 10 PeV, IceTopSMT starts losing efficiency at zenith angles above $\sim$ 54$^\circ$. In comparison, the IceTop7HG trigger continues to be fully efficient for zenith angles above $\sim$ 54$^\circ$ for 10 PeV showers. So, there is an improvement in trigger efficiency for inclined events above 10 PeV with the IceTop7HG trigger. In the future, when there will be several surface array enhancement stations with radio antennas, it is expected that with the implementation of the IceTop7HG trigger in IceTop, coincident detection of air showers with radio and IceTop tanks can be done for energies > 10 PeV where radio antennas start to get sensitive to air showers at the South Pole~\cite{IceCube:2023rrl}.





\section{Test with the IceTop7HG trigger at the South Pole}\label{sec:Test}
As the simulation showed that the IceTop7HG trigger was able to lower the energy threshold of full trigger efficiency for inclined air showers, the next step in the development of the trigger was to test if the trigger can be implemented in the data acquisition system of the detector. IceTop7HG was first tested at the South Pole Test System (SPTS), a test DAQ system similar to the system deployed for the IceCube detector to see if the trigger can be handled by the system. Recorded IceCube hits were rerun through the test DAQ that now included the IceTop7HG trigger in the configuration setting. The DAQ could handle the additional single tank hits in the input and produce events including those related to the IceTop7HG trigger without any issues.

Afterwards, a short 10 minute test run was conducted at the actual IceCube DAQ at the South Pole on June 6, 2023. We analyzed the events with the new trigger, IceTop7HG in the raw data produced from the short test run. The trigger rate  of the IceTop7HG trigger was found to be $\sim$ 31 Hz. As expected due to the additional snow accumulation on top of IceTop, this rate is lower than the IceTop7HG trigger rate found in the 2021 FRT data.

A longer 24 hour test run of the IceCube system was conducted over September $6\textup{--}7$, 2023 with the new configuration that included all planned changes to the trigger and filter settings, among them the IceTop7HG trigger. 

\begin{table}[htbp]
\centering
\caption{IceTop7HG trigger rate in 24 hr test run\label{tab:testRun}}
\smallskip
\begin{tabular}{|l|l|l|c|}
\hline
Run&Start time &Duration [hh:mm:ss]&Rate [Hz]\\
\hline
138329 & 2023-09-06 20:56:29&8:00:00 & 32.22\\
\hline
138330 & 2023-09-07 04:56:29&8:00:10 & 32.23\\
\hline
138331 & 2023-09-07 13:15:13&7:59:46 & 31.43\\
\hline
\end{tabular}
\end{table}

~\Cref{tab:testRun} shows the IceTop7HG trigger rate measured during three $\sim$ 8 hour runs of the 24 hour test run. The trigger rate is $\sim$ 32 Hz, consistent with what was measured in the shorter 10 minute test run. So, the IceCube DAQ is able to handle the new trigger, IceTop7HG without any significant issues. The IceTop7HG trigger was successfully implemented for the 2023/2024 data taking season of IceCube which started on November 28, 2023. In addition, there is an ongoing effort to design a filter that will select inclined cosmic-ray air shower events that can be reconstructed and used for physics analyses.

\section{Conclusion}
The new IceTop trigger, IceTop7HG, selects events where there are at least 7 tanks with coincident hits of high gain DOMs within 3 $\mu$s. Simulation showed that the IceTop7HG trigger stays fully efficient for cosmic-ray air showers with primary energy above 10 PeV even for inclination angles of 56.8$^\circ$ $\leq$ $\theta$ $\leq$ 65.0$^\circ$ which is an improvement over the standard IceTopSMT trigger in that phase space. Measured data with the Fixed Rate Trigger was used to get a realistic estimate of the total rate of the IceTop7HG trigger and the rate was found to be $\lesssim$ 40 Hz for the year 2021. This rate is expected to decrease further for the 2023 season due to additional years of snow accumulation and was $\sim$32 Hz in the 2023 test run. The test of the triggering system with the new trigger was done on the test DAQ system as well as on the actual DAQ of IceCube, first for 10 minute and then for a longer 24 hour test run. The trigger system was able to handle the higher rate of inputs that included additional single tank hits from IceTop on top of coincident station hits and produce events with the new trigger without any issue. A filter to select inclined events from the pool of triggered events is under development.

\acknowledgments
The authors gratefully acknowledge the support from the following agencies and institutions: USA – U.S. National Science Foundation-Office of Polar Programs, U.S. National Science Foundation-Physics Division, U.S. National Science Foundation-EPSCoR, U.S. National Science Foundation-Office of Advanced Cyberinfrastructure, Wisconsin Alumni Research Foundation, Center for High Throughput Computing (CHTC) at the University of Wisconsin–Madison, Open Science Grid (OSG), Partnership to Advance Throughput Computing (PATh), Advanced Cyberinfrastructure Coordination Ecosystem: Services \& Support (ACCESS), Frontera computing project at the Texas Advanced Computing Center, U.S. Department of Energy-National Energy Research Scientific Computing Center, Particle astrophysics research computing center at the University of Maryland, Institute for Cyber-Enabled Research at Michigan State University, Astroparticle physics computational facility at Marquette University, NVIDIA Corporation, and Google Cloud Platform; Belgium – Funds for Scientific Research (FRS-FNRS and FWO), FWO Odysseus and Big Science programmes, and Belgian Federal Science Policy Office (Belspo); Germany – Bundesministerium für Bildung und Forschung (BMBF), Deutsche Forschungsgemeinschaft (DFG), Helmholtz Alliance for Astroparticle Physics (HAP), Initiative and Networking Fund of the Helmholtz Association, Deutsches Elektronen Synchrotron (DESY), and High Performance Computing cluster of the RWTH Aachen; Sweden – Swedish Research Council, Swedish Polar Research Secretariat, Swedish National Infrastructure for Computing (SNIC), and Knut and Alice Wallenberg Foundation; European Union – EGI Advanced Computing for research; Australia – Australian Research Council; Canada – Natural Sciences and Engineering Research Council of Canada, Calcul Québec, Compute Ontario, Canada Foundation for Innovation, WestGrid, and Digital Research Alliance of Canada; Denmark – Villum Fonden, Carlsberg Foundation, and European Commission; New Zealand – Marsden Fund; Japan – Japan Society for Promotion of Science (JSPS) and Institute for Global Prominent Research (IGPR) of Chiba University; Korea – National Research Foundation of Korea (NRF); Switzerland – Swiss National Science Foundation (SNSF). 



\bibliographystyle{JHEP}
\bibliography{biblio.bib}






\end{document}